\documentclass[conference,a4paper]{APSIPA2020}
\usepackage{graphicx}
\usepackage{amsmath}
\usepackage{amssymb}
\usepackage{amsfonts}
\usepackage{amsxtra}
\usepackage{mathrsfs}
\usepackage{multirow}
\usepackage{threeparttable}
\usepackage{arydshln}
\usepackage{cite}
\usepackage{url}
\usepackage[keeplastbox]{flushend}
\usepackage{afterpage}

\makeatletter
\let\MYcaption\@makecaption
\makeatother
\usepackage{subcaption}
\makeatletter
\let\@makecaption\MYcaption
\makeatother
\newcommand{\myvector}[1]{\boldsymbol{#1}}
\newcommand{\mymatrix}[1]{\mathrm{#1}}
\newcommand{\mytensor}[1]{\boldsymbol{\mathrm{#1}}}

\newcounter{num}

\begin{document}

\title{Checkerboard-Artifact-Free Image-Enhancement Network
  Considering Local and Global Features}

\author{%
\authorblockN{Yuma Kinoshita and Hitoshi Kiya}%
\authorblockA{Tokyo Metropolitan University, Tokyo, Japan}%
}
\maketitle
\thispagestyle{empty}
\begin{abstract}
  In this paper, we propose a novel convolutional neural network (CNN)
  that never causes checkerboard artifacts, for image enhancement.
  In research fields of image-to-image translation problems,
  it is well-known that images generated by usual CNNs
  are distorted by checkerboard artifacts
  which mainly caused in forward-propagation of upsampling layers.
  However, checkerboard artifacts in image enhancement have never been discussed.
  In this paper, we point out that applying U-Net based CNNs to image enhancement
  causes checkerboard artifacts.
  In contrast, the proposed network that contains fixed convolutional layers
  can perfectly prevent the artifacts.
  In addition, the proposed network architecture, which can handle
  both local and global features,
  enables us to improve the performance of image enhancement.
  Experimental results show that the use of fixed convolutional layers
  can prevent checkerboard artifacts and the proposed network outperforms
  state-of-the-art CNN-based image-enhancement methods
  in terms of various objective quality metrics: PSNR, SSIM, and NIQE.
\end{abstract}
\section{Introduction}
  The low dynamic range of modern digital cameras
  is a major factor that prevents cameras from capturing images as well as human vision.
  This is due to the limited dynamic range that imaging sensors have,
  resulting in low-contrast images.
  Enhancing such images reveals hidden details.

  Various kinds of research on single-image enhancement have been reported
  \cite{zuiderveld1994contrast, wu2017contrast,
  guo2017lime, fu2016weighted, kinoshita2018automatic_trans}.
  Most image enhancement methods can be divided into two types:
  histogram equalization (HE)-based methods and Retinex-based methods.
  Additionally, multi-exposure-fusion (MEF)-based single-image enhancement methods
  have also been proposed
  ~\cite{kinoshita2019scene, kinoshita2018automatic_trans, ying2017bio}.
  However, these analysis-based methods cannot restore lost pixel values
  due to quantizing and clipping.
  The problem leads to banding artifacts in enhanced images.

  Recent work has demonstrated great progress by using
  convolutional neural networks (CNNs)
  in preference to analytical approaches such as HE
  ~\cite{gharbi2017deep, shen2017msrnet, cai2018learning, chen2018learning, yang2018image,
  ruixing2019underexposed, jiang2019enlighten, kinoshita2019convolutional}.
  Most of these methods employ a U-Net \cite{ronneberger2015unet}-based
  network architecture in order to learn the mapping from low-quality images
  to high-quality ones.
  However, it is well-known that CNNs having upsampling layers such as U-Net
  cause images to be distorted by checkerboard artifacts
  due to upsampling layers
  ~\cite{odena2016deconvolution, sugawara2018super,
  sugawara2019checkerboard, kinoshita2020fixed}.
  Despite this situation,
  checkerboard artifacts have never been discussed
  in the field of image enhancement so far.

  In this paper, we point out that applying U-Net based CNNs to image enhancement
  causes checkerboard artifacts.
  To prevent the artifacts, we also propose a novel image-enhancement CNN
  that never causes checkerboard artifacts.
  In the proposed network,
  fixed convolutional layers~\cite{kinoshita2020fixed} are applied
  to upsampling and downsampling operations.
  The use of fixed convolutional layers
  can perfectly prevent checkerboard artifacts.
  Moreover, the proposed network architecture that
  can handle both local and global features
  enables us to prevent distortions resulting from the lack of global image information.

  We evaluate the effectiveness of the proposed image-enhancement network
  in terms of the quality of enhanced images
  by an experiment using a dataset from~\cite{cai2018learning},
  where the peak signal-to-noise ratio (PSNR),
  the structural similarity (SSIM), and discrete entropy
  are utilized as quality metrics.
  Experimental results show that the proposed method outperforms
  state-of-the-art contrast enhancement methods
  in terms of those quality metrics.
  Furthermore, the proposed method does not cause checkerboard artifacts
  and distortions resulting from the lack of global features.
  
\section{Related work}
  Here, we briefly summarize image-enhancement methods
  and checkerboard artifacts in CNNs.

\subsection{Image enhancement}
  Many image-enhancement methods have been studied
  \cite{zuiderveld1994contrast, wu2017contrast,
  guo2017lime, fu2016weighted,
  kinoshita2018automatic_trans, kinoshita2019scene}.
  Among the methods, HE has received
  the most attention because of its intuitive implementation quality and
  high efficiency.
  It aims to derive a mapping function such that the entropy of
  a distribution of output luminance values can be maximized.
  However, HE often causes over-enhancement.
  To avoid this,
  numerous improved methods based on HE have also been developed
  \cite{zuiderveld1994contrast, wu2017contrast}.

  Another way for enhancing images is to use the Retinex theory \cite{land1977retinex}.
  Retinex-based methods \cite{guo2017lime, fu2016weighted} decompose images into
  reflectance and illumination, and then enhance images by manipulating illumination.
  Additionally, multi-exposure-fusion (MEF)-based single-image enhancement methods
  were recently proposed
  \cite{kinoshita2019scene, kinoshita2018automatic_trans, ying2017bio}.
  One of them, a pseudo MEF scheme~\cite{kinoshita2018automatic_trans},
  makes any single image applicable
  to MEF methods~\cite{mertens2009exposure}
  by generating pseudo multi-exposure images from a single image.
  By using this scheme, images with improved quality are produced
  with the use of detailed local features.
  
  Recent work has demonstrated great progress by using data-driven approaches
  in preference to traditional analytical approaches such as HE
  ~\cite{gharbi2017deep, shen2017msrnet, cai2018learning, chen2018learning, yang2018image,
  ruixing2019underexposed, jiang2019enlighten, kinoshita2019convolutional}.
  These data-driven approaches utilize high- and low-quality images
  to train deep neural networks,
  and the trained networks can be used to enhance color images.
  Most of these methods employ a U-Net \cite{ronneberger2015unet}-based
  network architecture.
  For example, Cai et al. utilized U-Net for enhancing a luminance map
  calculated from an input image~\cite{cai2018learning}.
  However, U-Net based network architectures cause checkerboard artifacts
  as described in the following section.

\subsection{Checkerboard artifacts in CNNs}
  Checkerboard artifacts have been studied as a distortion
  caused by using upsamplers in linear multi-rate systems
  ~\cite{harada1998multidimensional, tamura1998design, harada1998multidimensional_trans,
  iwai2010methods}.
  In research fields of image-to-image translation problems,
  e.g., image super-resolution,
  checkerboard artifacts are known to be
  caused by forward-propagation of upsampling layers
  including transposed convolutional layers
  and by backward-propagation of downsampling layers including strided convolutional layers
  ~\cite{odena2016deconvolution}.
  CNN architectures usually have upsampling layers and/or have downsampling layers,
  such as VGG~\cite{simonyan2014very}, ResNet~\cite{he2016deep},
  and U-Net~\cite{ronneberger2015unet},
  for increasing and/or reducing the spatial sampling rate of feature maps,
  respectively~\cite{goodfellow2016deep}.
  For this reason,
  checkerboard artifacts affect most commonly-used CNNs.
  In particular, the checkerboard artifacts caused by forward-propagation
  directory distort images generated by CNNs.

  To overcome checkerboard artifacts caused by upsampling layers,
  Sugawara et al.~\cite{sugawara2018super, sugawara2019checkerboard}
  gave us two approaches to perfectly prevent checkerboard artifacts
  by extending a condition for avoiding checkerboard artifacts
  in linear multirate systems
  ~\cite{harada1998multidimensional, tamura1998design, harada1998multidimensional_trans,
  iwai2010methods}.
  In accordance with Sugawara's approaches,
  Kinoshita et al. proposed fixed smooth convolutional layers
  that can prevent checkerboard artifacts caused by both upsampling
  and downsampling layers~\cite{kinoshita2020fixed}.

\subsection{Scenario}
  In image enhancement, 
  it has already been pointed out that
  CNNs that cannot handle global image information
  cause images to be distorted~\cite{marnerides2018expandnet, kinoshita2019convolutional}.
  In addition to this distortion,
  checkerboard artifacts will appear in enhanced images because of upsampling layers.

\begin{figure}[!t]
    \centering
    \begin{subfigure}[t]{0.45\hsize}
      \centering
      \includegraphics[width=\columnwidth]{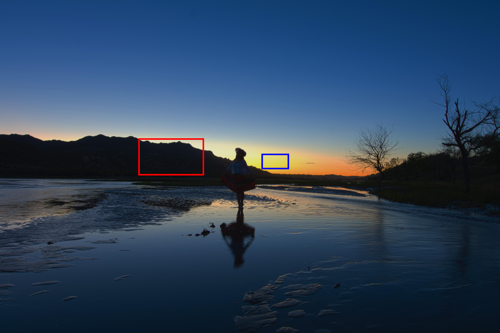}\\
      \vspace{2pt}
      \includegraphics[width=0.47\columnwidth]{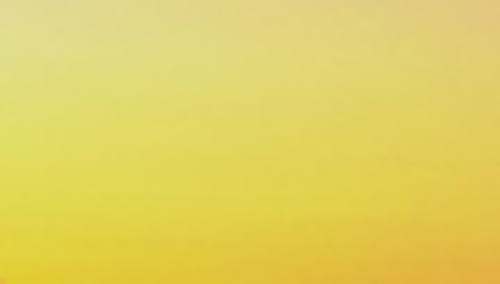}
      \includegraphics[width=0.47\columnwidth]{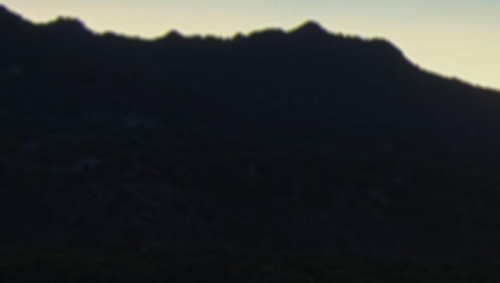}
      \caption{Input image \label{fig:input_example}}
    \end{subfigure}
    \begin{subfigure}[t]{0.45\hsize}
      \centering
      \includegraphics[width=\columnwidth]{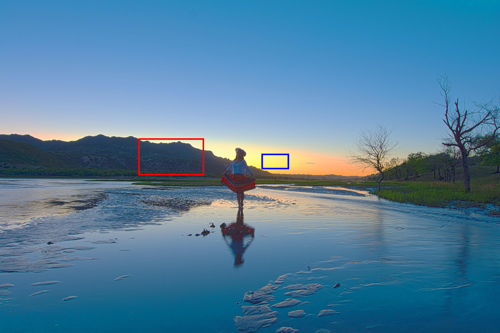}\\
      \vspace{2pt}
      \includegraphics[width=0.47\columnwidth]{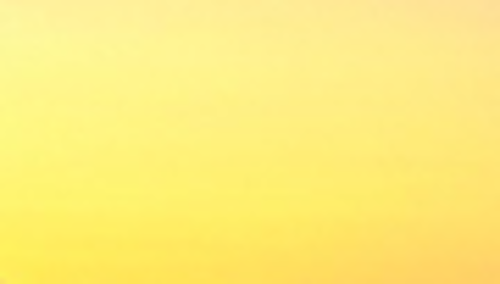}
      \includegraphics[width=0.47\columnwidth]{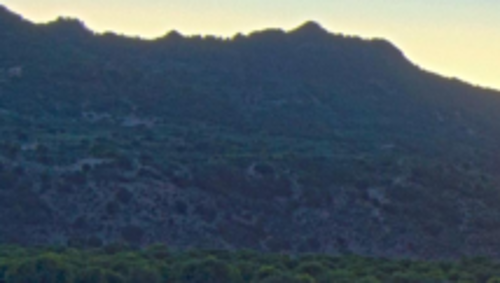}
      \caption{Ground truth \label{fig:target_example}}
    \end{subfigure}\\
    \begin{subfigure}[t]{0.45\hsize}
      \centering
      \includegraphics[width=\columnwidth]{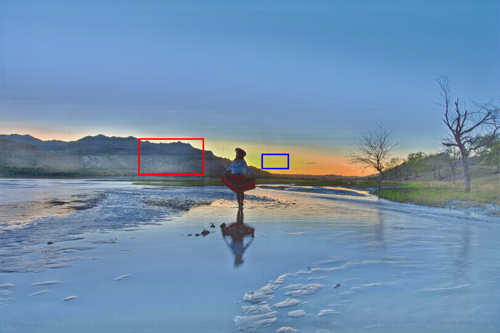}\\
      \vspace{2pt}
      \includegraphics[width=0.47\columnwidth]{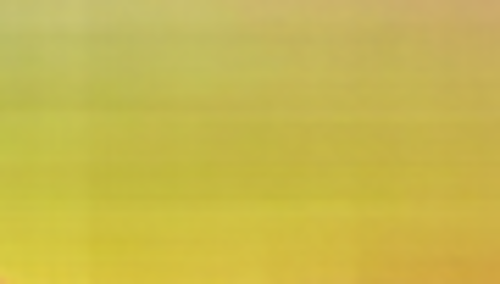}
      \includegraphics[width=0.47\columnwidth]{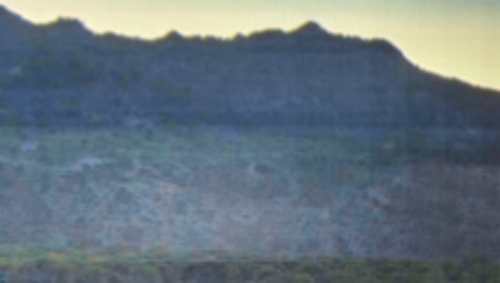}
      \caption{U-Net \label{fig:unet_example}}
    \end{subfigure}
    \begin{subfigure}[t]{0.45\hsize}
      \centering
      \includegraphics[width=\columnwidth]{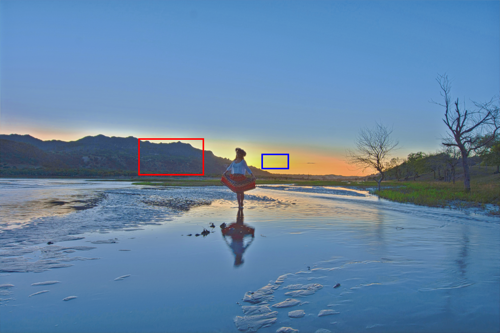}\\
      \vspace{2pt}
      \includegraphics[width=0.47\columnwidth]{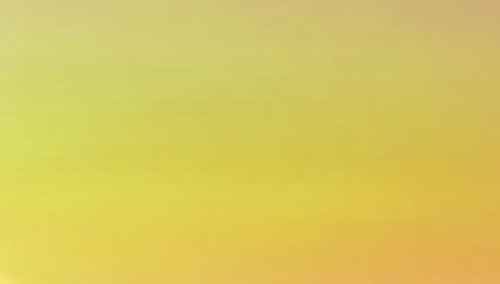}
      \includegraphics[width=0.47\columnwidth]{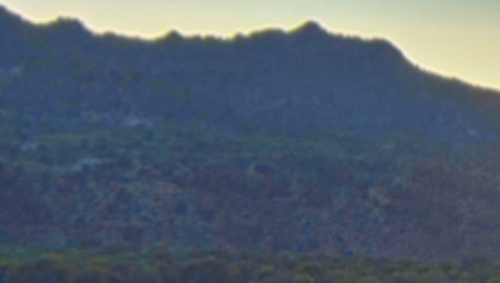}
      \caption{iTM-Net \label{fig:itm-net_example}}
    \end{subfigure}
    \caption{Checkerboard artifacts caused by upsampling layers
      and distortions due to the lack of global features.
      Zoom-ins of boxed regions are shown in bottom of each image.
      \label{fig:artifacts}}
  \end{figure}

  Figure~\ref{fig:artifacts} shows an example of images enhanced
  by using U-Net and iTM-Net~\cite{kinoshita2019convolutional},
  where iTM-Net can handle global features but U-Net cannot.
  From Fig.~\ref{fig:artifacts}(\subref{fig:unet_example}),
  there are luminance inversions in many parts of the image enhanced by using U-Net.
  These luminance inversions are distortions due to the lack of global features.
  In addition, checkerboard artifacts also appeared in the image.
  iTM-Net prevented the distortions but checkerboard artifacts still remain
  [see Fig.~\ref{fig:artifacts}(\subref{fig:itm-net_example})].

  For these reasons,
  CNN architectures for image enhancement should satisfy the following two conditions:
  \begin{itemize}
    \item Both local and global features of images can be handled.
    \item Checkerboard artifacts caused by upsampling layers can be prevented.
  \end{itemize}
  However, there are no CNNs that satisfy the conditions.
  Therefore, in this paper, we propose a novel image-enhancement network
  that enables us not only to perfectly prevent checkerboard artifacts
  but also to consider both local and global features.

\section{Proposed method}
\begin{figure*}[t]
    \centering
    \includegraphics[clip, width=0.95\hsize]{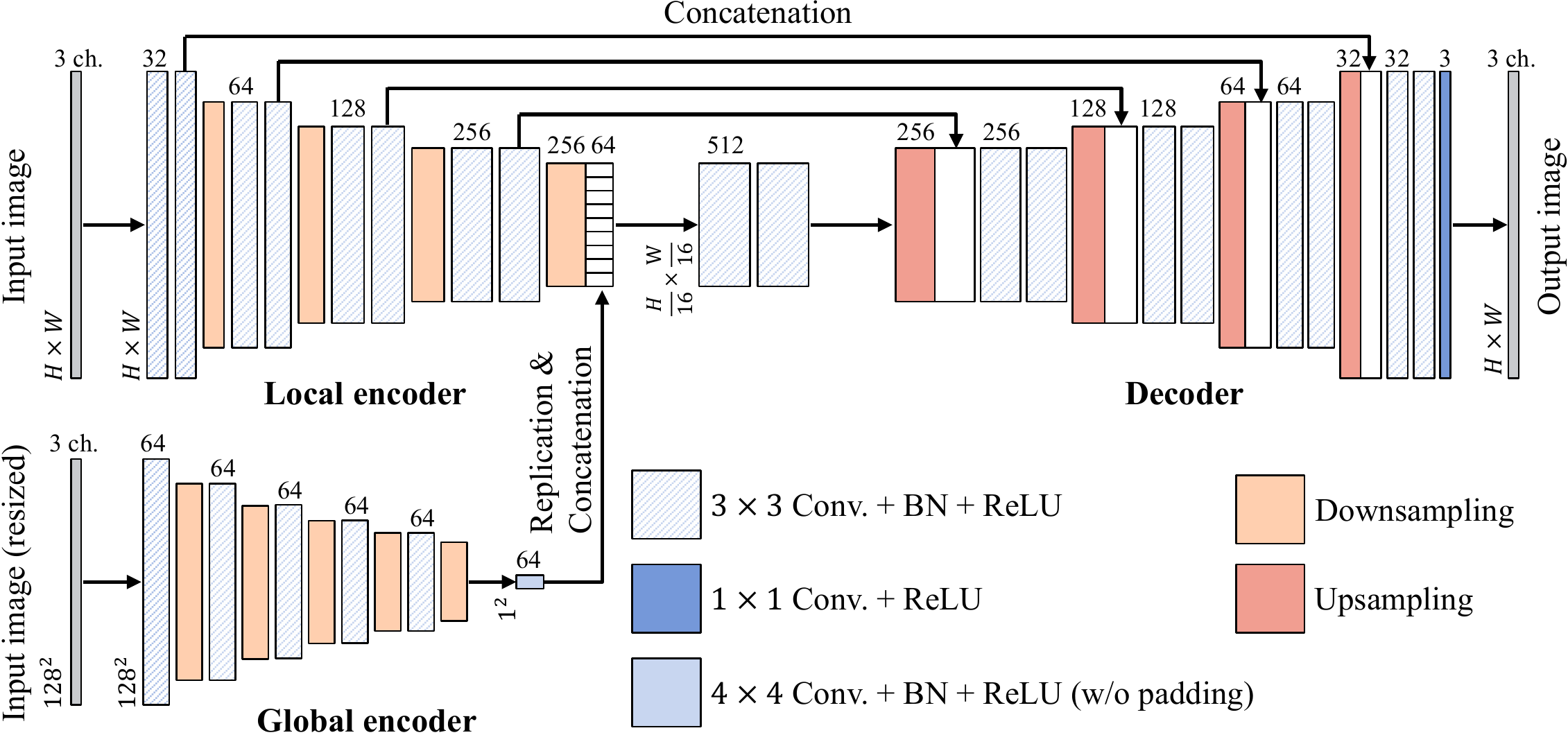}
    \caption{Network architecture.
      Architecture consists of local encoder, global encoder, and decoder.
      For upsampling and downsampling, fixed convolutional layers are applied
      to transposed convolution and strided convolution, respectively.
      Each box denotes multi-channel feature map produced by each layer.
      Number of channels is denoted above each box.
      Feature map resolutions are denoted to the left of a box.
      \label{fig:network}}
  \end{figure*}

  As shown in Fig. \ref{fig:network},
  the architecture of the proposed image-enhancement network
  consists of three sub-networks: a local encoder, a global encoder, and a decoder.
  The use of the local encoder and the global encoder makes it possible
  to handle local and global features, respectively.
  Furthermore, checkerboard artifacts in the network are perfectly prevented by using
  fixed convolutional layers~\cite{kinoshita2020fixed} in both upsampling and downsampling.

\subsection{Fixed convolutional layer}
  As in~\cite{sugawara2018super, sugawara2019checkerboard, kinoshita2020fixed},
  checkerboard artifacts can be perfectly prevented by inserting a fixed convolutional layer
  after each upsampling layer and before each downsampling layer.
  The filter kernel $\mytensor{K}^{(d)}$ of the fixed convolutional layer is obtained
  by convolving a zero-order hold kernel $\mymatrix{h}_0$ multiple times, as
  \begin{equation}
    \mytensor{K}^{(d)}_{i, i, :, :} =
    \begin{cases}
      \mymatrix{h}_0 & (d = 0) \\
      \mytensor{K}^{(d-1)}_{i, i, :, :} * \mymatrix{h}_0 & (d > 1)
    \end{cases},
    \label{eq:fixed_kernel}
  \end{equation}
  where a parameter $d$, referred to as the order of smoothness,
  $\mytensor{K}^{(d)}_{i, i, :, :}$ is 2-D slice of 4-D tensor $\mytensor{K}^{(d)}$,
  and $\mymatrix{A} * \mymatrix{B} $ means
  the convolution on two matrices $\mymatrix{A}$ and $\mymatrix{B}$.
  When the fixed layer is applied to an upsampling layer with an upsampling rate $U$,
  the kernel size for $\mymatrix{h}_0$ is given as $U \times U$.
  In contrast,
  the kernel size is given as $D \times D$
  when the proposed layer is applied to a downsampling layer with a downsampling rate $D$.

  By using a filter kernel $\mytensor{K}^{(d)}$ and a trainable bias $\myvector{b}$,
  an output feature map $\mytensor{Z}$ of the fixed layer
  can be written as
  \begin{equation}
    \mytensor{Z}_{i, j, k} = \sum_{m, n} \mytensor{V}_{i, j+m-1, k+n-1}
                                         \mytensor{K}^{(d)}_{i, i, m, n} + \myvector{b}_i,
  \end{equation}
  where $\myvector{V}$ is an input feature map with a size of
  \textit{channel} $\times$ \textit{height} $\times$ \textit{width}.

\subsection{Network architecture}
  Figure \ref{fig:network} shows the overall network architecture of
  the proposed network.
  The architecture consists of three sub-networks:
  a local encoder, a global encoder, and a decoder,
  which is based on iTM-Net used in
  ~\cite{kinoshita2019convolutional, kinoshita2019itmnet_trans}.
  iTM-Net utilizes the global encoder in addition to U-Net's architecture
  and combines features extracted by both encoders to prevent the distortions.
  In addition to iTM-Net,
  the use of the fixed convolutional layers for upsampling and downsampling
  enables us to prevent checkerboard artifacts.

  The input for the local encoder is an $H \times W$ pixels 24-bit color image.
  For the global encoder,
  the input image is resized to a fixed size ($128 \times 128$).

  The proposed network have five types of layers as shown in Fig. \ref{fig:network}:
  \subsubsection*{$3 \times 3$ Conv. + BN + ReLU}
    which calculates a $3 \times 3$ convolution
    with a stride of $1$ and a padding of $1$.
    After the convolution, batch normalization \cite{ioffe2015batch}
    and the rectified linear unit activation function \cite{glorot2011deep} (ReLU)
    are applied.
    In the local encoder and the decoder, two adjacent 
    $3 \times 3$ Conv. + BN + ReLU layers will have the same number $K$ of filters.
    From the first two layers to the last ones,
    the number of filters are
    $K = 32$, $64$, $128$, $256$, $512$, $256$, $128$, $64$, and $32$,
    respectively.
    In the global encoder, all layers have $64$ filters.
  \subsubsection*{$1 \times 1$ Conv. + ReLU}
    which calculates a $1 \times 1$ convolution with a stride of $1$ and without padding.
    After the convolution, ReLU is applied.
    The number of filters in the layer is $3$.
  \subsubsection*{$4 \times 4$ Conv. + BN + ReLU (w/o padding)}
    which calculates a $4 \times 4$ convolution without padding.
    The number of filters in the layer is $64$.
  \subsubsection*{Downsampling}
    which calculates a $2 \times 2$ convolution with a stride of $1$ and a padding of $1$,
    by using a fixed kernel in Eq. (\ref{eq:fixed_kernel}).
    After the convolution, it downsamples feature maps by a $3 \times 3$ strided convolution
    with a stride of $2$ and a padding of $1$.
  \subsubsection*{Upsampling}
    which upsamples feature maps by a $4 \times 4$ transposed convolution
    with a stride of $1/2$ and a padding of $1$.
    After the transposed convolution,
    it calculates a $2 \times 2$ convolution with a stride of $1$ and a padding of $1$,
    by using a fixed kernel in Eq. (\ref{eq:fixed_kernel}).

\section{Simulation}
  We evaluated the effectiveness of the proposed method
  by using three objective quality metrics.

\subsection{Simulation conditions}
  In our experiments, we trained the proposed network with 100 epochs
  by using training data in a multi-exposure image dataset
  which which constructed by Cai et al.~\cite{cai2018learning}.

  For data augmentation,
  we resized each original input image with a random scaling factor in the range $[0.6, 1.0]$
  for every epoch.
  After the resizing, we randomly cropped the resized image to
  an image patch with a size of $256 \times 256$ pixels
  and flipped the patch horizontally with a probability of 0.5.

  Loss between an image patch generated by the proposed network
  and the corresponding target image patch was calculated by the simple $\ell 1$-distance.
  Here, the Adam optimizer \cite{kingma2014adam} was utilized for optimization,
  where parameters in Adam were set as $\alpha=0.001, \beta_1=0.9$, and $\beta_2=0.999$.
  He's method \cite{he2015delving} was used for initializing the network.

  As in~\cite{cai2018learning},
  we tested the proposed network by using underexposed and overexposed images
  having -1 and 1 $\mathrm{[EV]}$, respectively, in test data in the dataset.
  The quality of images enhanced by the proposed network
  was evaluated by three objective quality metrics:
  the peak signal-to-noise ratio (PSNR),
  the structural similarity (SSIM)~\cite{wang2004image},
  and the naturalness image quality evaluator (NIQE) \cite{mittal2013making},
  where the target image corresponding to the input image utilized
  as a reference for PSNR and SSIM.

\subsection{Results}
  To confirm the effectiveness of the proposed network architecture,
  we first compare four network architectures:
  U-Net~\cite{ronneberger2015unet},
  U-Net with fixed convolutional layers,
  iTM-Net~\cite{kinoshita2019convolutional, kinoshita2019itmnet_trans},
  and iTM-Net with fixed convolutional layers (i.e., the proposed one).
  Please note that these networks were trained by using the same data and parameters
  as the proposed network, and the difference among the networks
  was only the network architecture.

  Table \ref{tab:ablation} illustrates the average scores of the objective assessment
  for 58 underexposed images and 58 overexposed ones,
  in terms of PSNR, SSIM, and NIQE.
  In the case of PSNR and SSIM,
  a larger value means a higher similarity between an enhanced image
  and a reference image.
  By contrast, a smaller value for NIQE indicates that an enhanced image has
  less distortions such as noise or blur.
  As shown in Table \ref{tab:ablation},
  the proposed method provided the highest average scores in the four networks
  for both underexposed and overexposed images, under all three metrics.
  \begin{table*}[!t]
    \centering
    \caption{Average scores of objective quality metrics for ablation study}
    \begin{tabular}{l|ccc|ccc} \hline \hline
      \multirow{2}{*}{Architecture} & \multicolumn{3}{c}{Underexposure}
        & \multicolumn{3}{|c}{Overexposure} \\
                            & PSNR  & SSIM            & NIQE  & PSNR  & SSIM   & NIQE  \\ \hline
      U-Net~\cite{ronneberger2015unet} & 16.92 & 0.8213 & 2.613
        & 16.69 & 0.7961 & 2.681 \\
      U-Net w/ fixed layers & 16.29 & 0.8152 & 2.475
        & 16.18 & 0.7912 & 2.559 \\
      iTM-Net~\cite{kinoshita2019convolutional} & 20.20 & \textbf{0.8575} & 2.584
        & 20.46 & 0.8433 & 2.633 \\
      iTM-Net w/ fixed layers (Proposed) & \textbf{20.52} & \textbf{0.8575} & \textbf{2.382}
        & \textbf{21.00} & \textbf{0.8455} & \textbf{2.496} \\ \hline \hline
    \end{tabular}
    \label{tab:ablation}
  \end{table*}

\begin{figure*}[!t]
    \centering
    \begin{subfigure}[t]{0.19\hsize}
      \centering
      \includegraphics[width=\columnwidth]{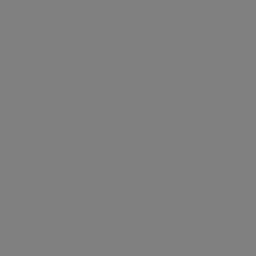} \\
      \vspace{1pt}
      \includegraphics[width=\columnwidth]{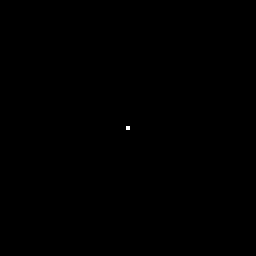}
      \caption{Input image \label{fig:input_ablation}}
    \end{subfigure}
    \begin{subfigure}[t]{0.19\hsize}
      \centering
      \includegraphics[width=\columnwidth]{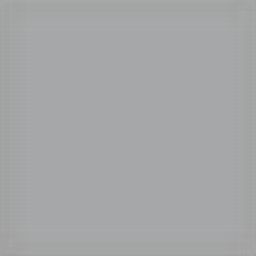} \\
      \vspace{1pt}
      \includegraphics[width=\columnwidth]{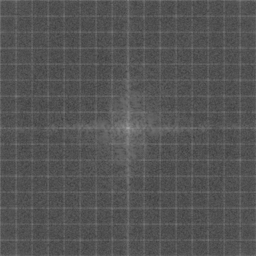}
      \caption{U-Net~\cite{ronneberger2015unet} \label{fig:unet_ablation}}
    \end{subfigure}
    \begin{subfigure}[t]{0.19\hsize}
      \centering
      \includegraphics[width=\columnwidth]{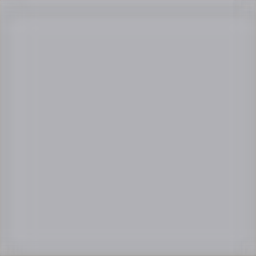} \\
      \vspace{1pt}
      \includegraphics[width=\columnwidth]{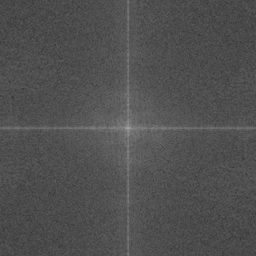}
      \caption{U-Net w/ fixed layers \label{fig:unet_wo_checker_ablation}}
    \end{subfigure}
    \begin{subfigure}[t]{0.19\hsize}
      \centering
      \includegraphics[width=\columnwidth]{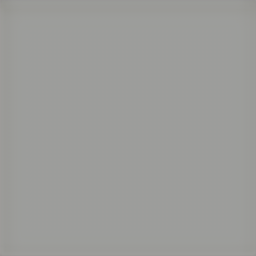} \\
      \vspace{1pt}
      \includegraphics[width=\columnwidth]{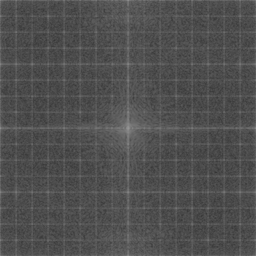}
      \caption{iTM-Net~\cite{kinoshita2019convolutional} \label{fig:itm-net_ablation}}
    \end{subfigure}
    \begin{subfigure}[t]{0.19\hsize}
      \centering
      \includegraphics[width=\columnwidth]{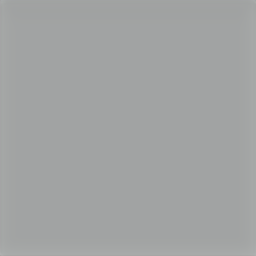} \\
      \vspace{1pt}
      \includegraphics[width=\columnwidth]{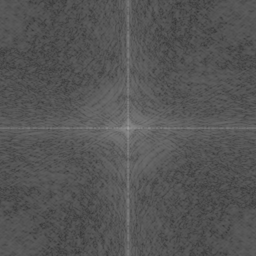}
      \caption{iTM-Net w/ fixed layers (Proposed) \label{fig:itm-net_wo_checker_ablation}}
    \end{subfigure}
    \caption{Checkerboard artifacts in generated images.
      Top: generated images. Bottom: log-amplitude spectra normalized in range $[0, 1]$.
      \label{fig:ablation}}
  \end{figure*}

  Figure \ref{fig:ablation} shows
  images generated from an artificial gray image by the four networks
  and their log-amplitude spectra,
  where each spectrum was normalized in the range $[0, 1]$.
  From the figure,
  it is confirmed that network architectures without fixed convolutional layers
  caused checkerboard artifacts in the generated images.
  These artifacts can also be seen as a lattice pattern in the frequency domain.
  In addition, the use of fixed convolutional layers
  perfectly prevented the artifacts.
  For these reasons, the architecture of the proposed network that can handle both
  local and global features and can prevent checkerboard artifacts
  is effective for image enhancement.

  The proposed network was also compared with seven conventional methods:
  histogram equalization (HE),
  contrast-accumulated histogram equalization (CACHE) \cite{wu2017contrast},
  simultaneous reflectance and illumination estimation (SRIE) \cite{fu2016weighted},
  low-light image enhancement via illumination map estimation (LIME) \cite{guo2017lime},
  pseudo multi-exposure image fusion (PMEF)~\cite{kinoshita2018automatic},
  deep underexposed photo enhancement (DeepUPE)~\cite{ruixing2019underexposed},
  and EnlightenGAN~\cite{jiang2019enlighten},
  where HE and CACHE are HE-based methods,
  SRIE and LIME are Retinex-based ones,
  PMEF is an MEF-based one,
  and DeepUPE and EnlightenGAN are deep-learning-based ones.

\begin{figure*}[!t]
    \centering
    \begin{subfigure}[t]{0.19\hsize}
      \centering
      \includegraphics[width=\columnwidth]{./figs/input_under.png}\\
      \vspace{2pt}
      \includegraphics[width=0.47\columnwidth]{./figs/input_under_zoom1.png}
      \includegraphics[width=0.47\columnwidth]{./figs/input_under_zoom2.png}
      \caption{Input image \label{fig:input_under}}
    \end{subfigure}
    \begin{subfigure}[t]{0.19\hsize}
      \centering
      \includegraphics[width=\columnwidth]{./figs/target_under.png}\\
      \vspace{2pt}
      \includegraphics[width=0.47\columnwidth]{./figs/target_under_zoom1.png}
      \includegraphics[width=0.47\columnwidth]{./figs/target_under_zoom2.png}
      \caption{Ground truth \label{fig:target_under}}
    \end{subfigure}
    \begin{subfigure}[t]{0.19\hsize}
      \centering
      \includegraphics[width=\columnwidth]{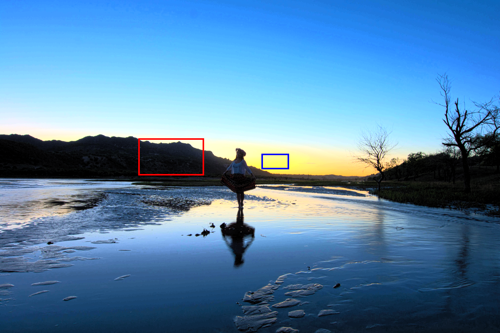}\\
      \vspace{2pt}
      \includegraphics[width=0.47\columnwidth]{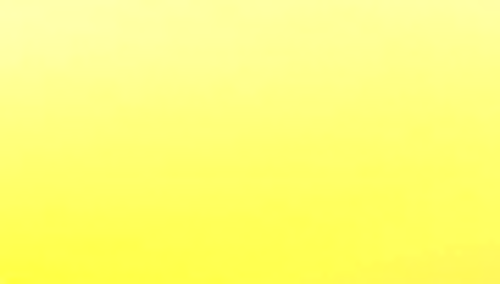}
      \includegraphics[width=0.47\columnwidth]{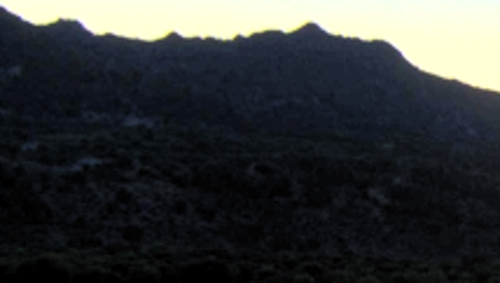}
      \caption{HE \label{fig:he_under}}
    \end{subfigure}
    \begin{subfigure}[t]{0.19\hsize}
      \centering
      \includegraphics[width=\columnwidth]{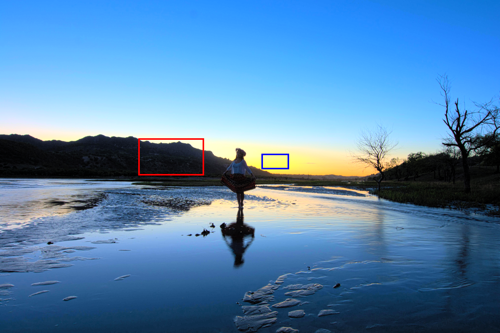}\\
      \includegraphics[width=0.47\columnwidth]{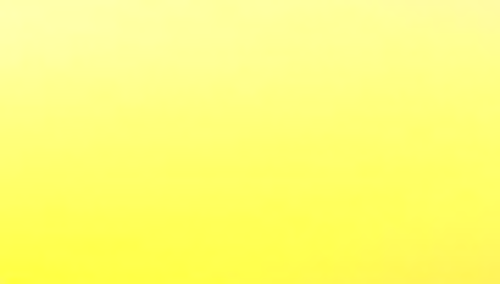}
      \includegraphics[width=0.47\columnwidth]{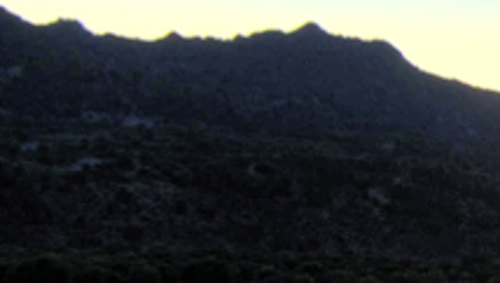}
      \caption{CACHE~\cite{wu2017contrast} \label{fig:cache_under}}
    \end{subfigure}
    \begin{subfigure}[t]{0.19\hsize}
      \centering
      \includegraphics[width=\columnwidth]{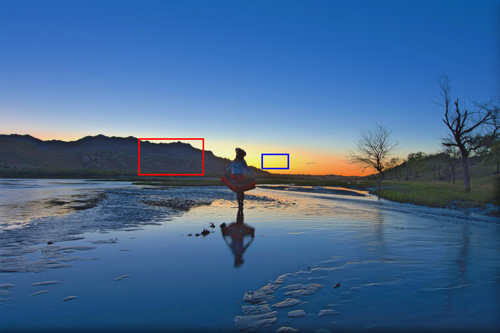}\\
      \vspace{2pt}
      \includegraphics[width=0.47\columnwidth]{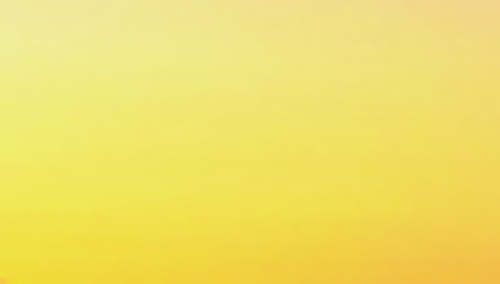}
      \includegraphics[width=0.47\columnwidth]{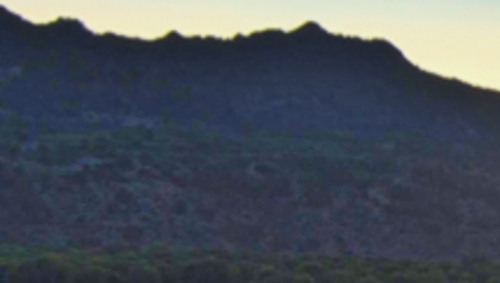}
      \caption{SRIE~\cite{fu2016weighted} \label{fig:srie_under}}
    \end{subfigure} \\
    \begin{subfigure}[t]{0.19\hsize}
      \centering
      \includegraphics[width=\columnwidth]{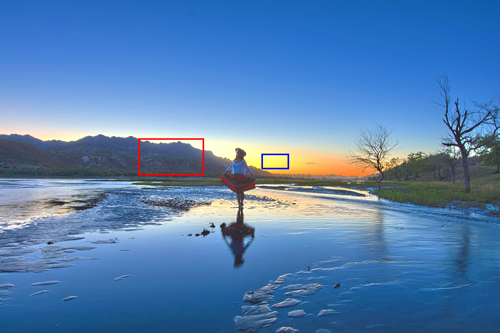}\\
      \vspace{2pt}
      \includegraphics[width=0.47\columnwidth]{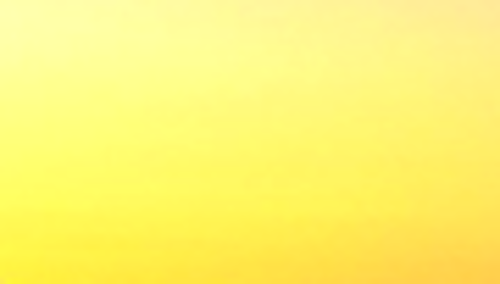}
      \includegraphics[width=0.47\columnwidth]{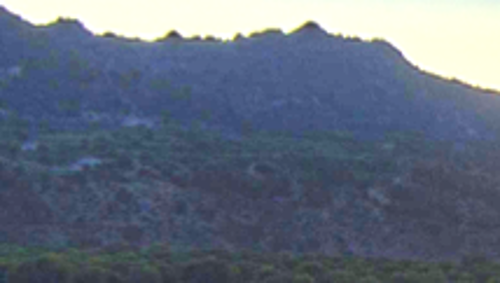}
      \caption{LIME~\cite{guo2017lime} \label{fig:lime_under}}
    \end{subfigure}
    \begin{subfigure}[t]{0.19\hsize}
      \centering
      \includegraphics[width=\columnwidth]{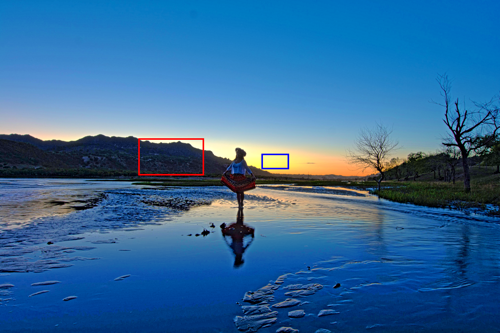}\\
      \vspace{2pt}
      \includegraphics[width=0.47\columnwidth]{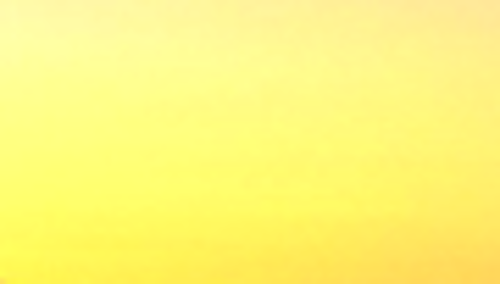}
      \includegraphics[width=0.47\columnwidth]{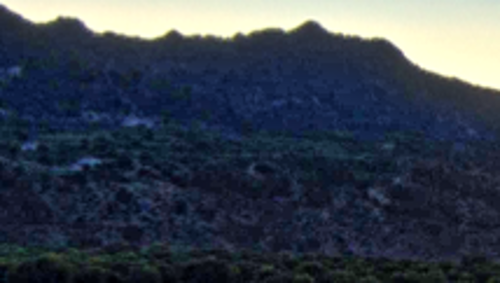}
      \caption{PMEF~\cite{kinoshita2018automatic} \label{fig:pmef_under}}
    \end{subfigure}
    \begin{subfigure}[t]{0.19\hsize}
      \centering
      \includegraphics[width=\columnwidth]{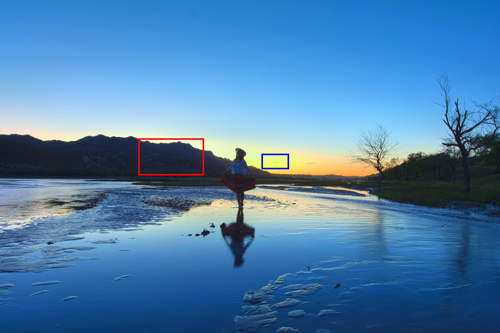}\\
      \vspace{2pt}
      \includegraphics[width=0.47\columnwidth]{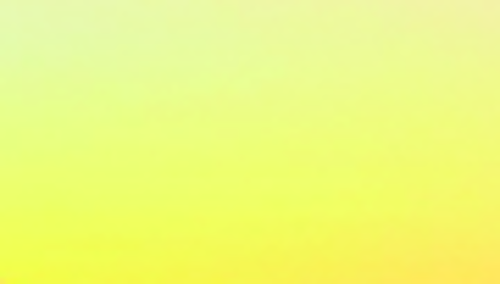}
      \includegraphics[width=0.47\columnwidth]{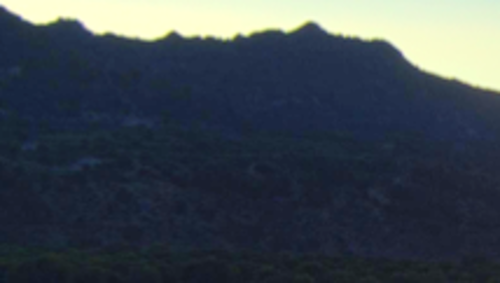}
      \caption{DeepUPE~\cite{ruixing2019underexposed} \label{fig:DeepUPE_under}}
    \end{subfigure}
    \begin{subfigure}[t]{0.19\hsize}
      \centering
      \includegraphics[width=\columnwidth]{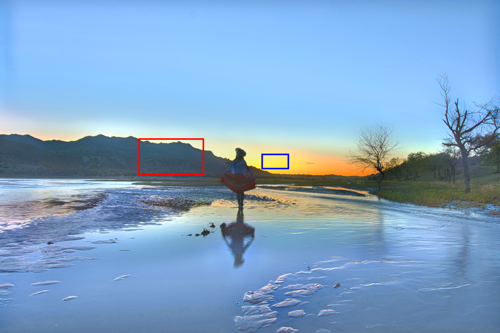}\\
      \vspace{2pt}
      \includegraphics[width=0.47\columnwidth]{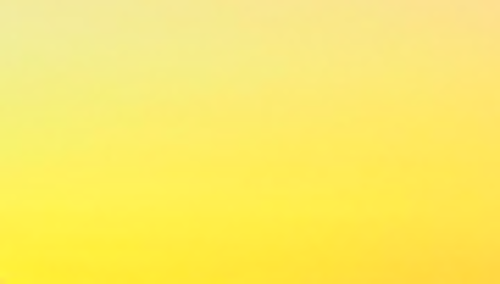}
      \includegraphics[width=0.47\columnwidth]{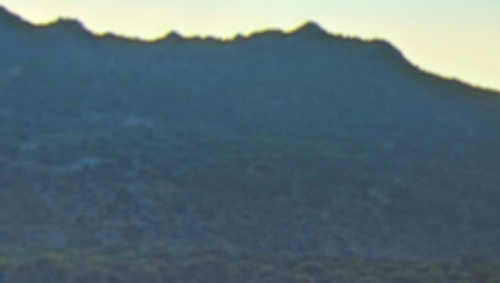}
      \caption{EnlightenGAN~\cite{jiang2019enlighten} \label{fig:EnlightenGAN_under}}
    \end{subfigure}
    \begin{subfigure}[t]{0.19\hsize}
      \centering
      \includegraphics[width=\columnwidth]{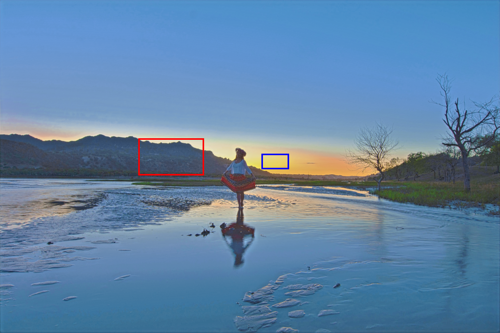}\\
      \vspace{2pt}
      \includegraphics[width=0.47\columnwidth]{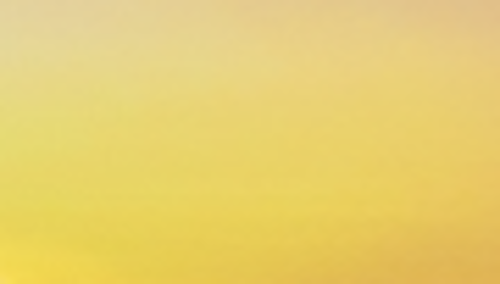}
      \includegraphics[width=0.47\columnwidth]{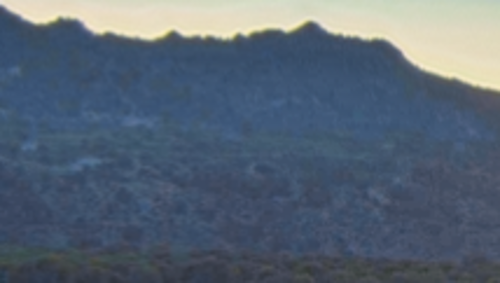}
      \caption{Proposed \label{fig:itm-net_wo_checker_under}}
    \end{subfigure}
    \caption{Example of enhanced images from underexposed image.
      Zoom-ins of boxed regions are shown in bottom of each image.
      \label{fig:comparison}}
  \end{figure*}

  Figure~\ref{fig:comparison} shows an example of images enhanced by the eight methods.
  From the figure, deep-learning-based methods were demonstrated to provide
  higher-quality images than conventional HE-, Retinex-, and MEF-based methods.
  In particular, the proposed network and EnlightenGAN generated
  images that clearly showed whole regions in the images.

  Table~\ref{tab:comparison} illustrates the average scores of the objective assessment
  for 58 underexposed images,
  in terms of PSNR, SSIM, and NIQE.
  Since compared methods aim to enhance low-light images,
  we did not evaluate scores for overexposed images.
  As shown in Table~\ref{tab:comparison},
  the proposed network provided the highest PSNR and SSIM scores in the eight methods.
  In contrast, conventional HE-, Retinex-, MEF-based methods produced 
  better NIQE scores than deep-learning-based methods,
  because of the simplicity of these traditional analysis-based methods.
  However,
  the proposed network provided the lowest NIQE score in the three deep-learning-based methods.
  Hence, it is confirmed that the proposed network architecture
  is unlikely to cause distortions.
  \begin{table*}[!t]
    \centering
    \caption{Average scores of objective quality metrics}
    \begin{tabular}{l|cccccccc} \hline \hline
      Metric & HE     & CACHE~\cite{wu2017contrast} & SRIE~\cite{fu2016weighted}
        & LIME~\cite{guo2017lime} & PMEF~\cite{kinoshita2018automatic}
        & DeepUPE~\cite{ruixing2019underexposed}
        & EnlightenGAN~\cite{jiang2019enlighten} & Proposed        \\ \hline
      PSNR   & 17.18  & 16.30          & 17.05  & 16.47  & 16.52
        & 16.59   & 15.83        & \textbf{20.52}  \\
      SSIM   & 0.7359 & 0.7521         & 0.7983 & 0.7991 & 0.7296
        & 0.7011  & 0.7677       & \textbf{0.8575} \\
      NIQE   & 2.768  & \textbf{2.268} & 2.524  & 2.345  & 2.309
        & 2.474   & 2.494        & 2.382           \\ \hline \hline
    \end{tabular}
    \label{tab:comparison}
  \end{table*}

  Those experimental results show that
  the proposed network architecture is effective for enhancing single images.

\section{Conclusion}
  In this paper, we proposed a novel image-enhancement network
  that never causes checkerboard artifacts.
  In the proposed network,
  the use of fixed convolutional layers perfectly prevents
  checkerboard artifacts caused by upsampling and downsampling.
  Furthermore,
  the architecture of the proposed network, which
  consists of a local encoder, a global encoder, and a decoder,
  can effectively prevent distortions
  resulting from the lack of global image information required for image enhancement.
  Experimental results showed that
  the use of the fixed layers enabled us to prevent checkerboard artifacts.
  In addition, the proposed network was shown to outperform
  state-of-the-art CNN-based image-enhancement methods,
  in terms of three objective quality metrics: PSNR, SSIM, and NIQE.
  
  In future work, we will train an image enhancement network
  with the proposed architecture by using more advanced loss functions,
  such as GAN-based ones,
  for improving the enhancement performance.

\section*{Acknowledgment}
  This work was supported by JSPS KAKENHI Grant Number JP18J20326.

%\bibliographystyle{IEEEtran}
%\bibliography{../../../../../bibliography/english_papers,../../../../../bibliography/standards,../../../../../bibliography/web_pages}

\begin{thebibliography}{10}
\providecommand{\url}[1]{#1}
\csname url@samestyle\endcsname
\providecommand{\newblock}{\relax}
\providecommand{\bibinfo}[2]{#2}
\providecommand{\BIBentrySTDinterwordspacing}{\spaceskip=0pt\relax}
\providecommand{\BIBentryALTinterwordstretchfactor}{4}
\providecommand{\BIBentryALTinterwordspacing}{\spaceskip=\fontdimen2\font plus
\BIBentryALTinterwordstretchfactor\fontdimen3\font minus
  \fontdimen4\font\relax}
\providecommand{\BIBforeignlanguage}[2]{{%
\expandafter\ifx\csname l@#1\endcsname\relax
\typeout{** WARNING: IEEEtran.bst: No hyphenation pattern has been}%
\typeout{** loaded for the language `#1'. Using the pattern for}%
\typeout{** the default language instead.}%
\else
\language=\csname l@#1\endcsname
\fi
#2}}
\providecommand{\BIBdecl}{\relax}
\BIBdecl

\bibitem{zuiderveld1994contrast}
K.~Zuiderveld, ``{Contrast Limited Adaptive Histogram Equalization},'' in
  \emph{Graph. gems IV}, P.~S. Heckbert, Ed.,
  San Diego, CA: Elsevier, 1994, pp. 474--485.

\bibitem{wu2017contrast}
\BIBentryALTinterwordspacing
X.~Wu, X.~Liu, K.~Hiramatsu, and K.~Kashino, ``{Contrast-accumulated histogram
  equalization for image enhancement},'' in \emph{Proc. IEEE Int. Conf. Image
  Process.}, Sep. 2017, pp. 3190--3194.
\BIBentrySTDinterwordspacing

\bibitem{guo2017lime}
\BIBentryALTinterwordspacing
X.~Guo, Y.~Li, and H.~Ling, ``{LIME: Low-Light Image Enhancement via
  Illumination Map Estimation},'' \emph{IEEE Trans. Image Process.}, vol.~26,
  no.~2, pp. 982--993, Feb. 2017.
\BIBentrySTDinterwordspacing

\bibitem{fu2016weighted}
\BIBentryALTinterwordspacing
X.~Fu, D.~Zeng, Y.~Huang, X.-P. Zhang, and X.~Ding, ``{A Weighted Variational
  Model for Simultaneous Reflectance and Illumination Estimation},'' in
  \emph{Proc. IEEE Conf. Comput. Vis. Pattern Recognit.},
  Jun. 2016, pp. 2782--2790.
\BIBentrySTDinterwordspacing

\bibitem{kinoshita2018automatic_trans}
\BIBentryALTinterwordspacing
Y.~Kinoshita and H.~Kiya, ``{Automatic exposure compensation using an image
  segmentation method for single-image-based multi-exposure fusion},''
  \emph{APSIPA Trans. Signal Inf. Process.}, vol.~7, e22, Dec. 2018.
\BIBentrySTDinterwordspacing

\bibitem{kinoshita2019scene}
\BIBentryALTinterwordspacing
------, ``{Scene Segmentation-Based Luminance Adjustment for Multi-Exposure
  Image Fusion},'' \emph{IEEE Trans. Image Process.}, vol.~28, no.~8, pp.
  4101--4116, Aug. 2019.
\BIBentrySTDinterwordspacing

\bibitem{ying2017bio}
\BIBentryALTinterwordspacing
Z.~Ying, G.~Li, and W.~Gao, ``{A Bio-Inspired Multi-Exposure Fusion Framework
  for Low-light Image Enhancement},'' \emph{arXiv: 1711.00591}, Nov.
  2017. [Online]. Available: \url{http://arxiv.org/abs/1711.00591}
\BIBentrySTDinterwordspacing

\bibitem{gharbi2017deep}
\BIBentryALTinterwordspacing
M.~Gharbi, J.~Chen, J.~T. Barron, S.~W. Hasinoff, and F.~Durand, ``{Deep
  bilateral learning for real-time image enhancement},'' \emph{ACM Trans.
  Graph.}, vol.~36, no.~4, pp. 1--12, Jul. 2017.
\BIBentrySTDinterwordspacing

\bibitem{shen2017msrnet}
\BIBentryALTinterwordspacing
L.~Shen, Z.~Yue, F.~Feng, Q.~Chen, S.~Liu, and J.~Ma, ``{MSR-net:Low-light
  Image Enhancement Using Deep Convolutional Network},'' \emph{arXiv: 1711.02488},
  Nov. 2017. [Online]. Available: \url{http://arxiv.org/abs/1711.02488}
\BIBentrySTDinterwordspacing

\bibitem{cai2018learning}
\BIBentryALTinterwordspacing
J.~Cai, S.~Gu, and L.~Zhang, ``{Learning a Deep Single Image Contrast Enhancer
  from Multi-Exposure Images},'' \emph{IEEE Trans. Image Process.}, vol.~27,
  no.~4, pp. 2049--2062, Apr. 2018.
\BIBentrySTDinterwordspacing

\bibitem{chen2018learning}
\BIBentryALTinterwordspacing
C.~Chen, Q.~Chen, J.~Xu, and V.~Koltun, ``{Learning to See in the Dark},'' in
  \emph{Proc. IEEE Conf. Comput. Vis. Pattern Recognit.},
  Jun. 2018, pp. 3291--3300.
\BIBentrySTDinterwordspacing

\bibitem{yang2018image}
\BIBentryALTinterwordspacing
X.~Yang, K.~Xu, Y.~Song, Q.~Zhang, X.~Wei, and R.~W. Lau, ``{Image Correction
  via Deep Reciprocating HDR Transformation},'' in \emph{Proc. IEEE Conf.
  Comput. Vis. Pattern Recognit.},
  Jun. 2018, pp. 1798--1807.
\BIBentrySTDinterwordspacing

\bibitem{ruixing2019underexposed}
\BIBentryALTinterwordspacing
R.~Wang, Q.~Zhang, C.-W. Fu, X.~Shen, W.-S. Zheng, and J.~Jia, ``{Underexposed
  Photo Enhancement Using Deep Illumination Estimation},'' in \emph{Proc. Conf.
  Comput. Vis. Pattern Recognit.},
  Jun. 2019, pp. 6842--6850.
\BIBentrySTDinterwordspacing

\bibitem{jiang2019enlighten}
\BIBentryALTinterwordspacing
Y.~Jiang, X.~Gong, D.~Liu, Y.~Cheng, C.~Fang, X.~Shen, J.~Yang, P.~Zhou, and
  Z.~Wang, ``{EnlightenGAN: Deep Light Enhancement without Paired
  Supervision},'' \emph{arXiv: 1906.06972}, Jun. 2019. [Online]. Available:
  \url{http://arxiv.org/abs/1906.06972}
\BIBentrySTDinterwordspacing

\bibitem{kinoshita2019convolutional}
\BIBentryALTinterwordspacing
Y.~Kinoshita and H.~Kiya, ``{Convolutional Neural Networks Considering Local
  and Global Features for Image Enhancement},'' in \emph{Proc. IEEE Int. Conf.
  Image Process.},
  Sep. 2019, pp. 2110--2114.
\BIBentrySTDinterwordspacing

\bibitem{ronneberger2015unet}
\BIBentryALTinterwordspacing
O.~Ronneberger, P.Fischer, and T.~Brox, ``{U-Net: Convolutional Networks for
  Biomedical Image Segmentation},'' in \emph{Med. Image Comput. Comput.
  Interv.}, ser. LNCS, vol. 9351,
  Springer, Nov. 2015, pp. 234--241.
\BIBentrySTDinterwordspacing

\bibitem{odena2016deconvolution}
\BIBentryALTinterwordspacing
A.~Odena, V.~Dumoulin, and C.~Olah, ``{Deconvolution and Checkerboard
  Artifacts},'' 2016. [Online]. Available:
  \url{https://distill.pub/2016/deconv-checkerboard/}
\BIBentrySTDinterwordspacing

\bibitem{sugawara2018super}
\BIBentryALTinterwordspacing
Y.~Sugawara, S.~Shiota, and H.~Kiya, ``{Super-Resolution Using Convolutional
  Neural Networks Without Any Checkerboard Artifacts},'' in \emph{Proc. IEEE
  Int. Conf. Image Process.},
  Oct. 2018, pp. 66--70.
\BIBentrySTDinterwordspacing

\bibitem{sugawara2019checkerboard}
\BIBentryALTinterwordspacing
------, ``{Checkerboard artifacts free convolutional neural networks},''
  \emph{APSIPA Trans. Signal Inf. Process.}, vol.~8, e9, Feb. 2019.
\BIBentrySTDinterwordspacing

\bibitem{kinoshita2020fixed}
\BIBentryALTinterwordspacing
Y.~Kinoshita and H.~Kiya, ``{Fixed Smooth Convolutional Layer for Avoiding
  Checkerboard Artifacts in CNNs},'' in \emph{Proc. IEEE Int. Conf. Acoust.
  Speech Signal Process.},
  May 2020, pp. 3712--3716.
\BIBentrySTDinterwordspacing

\bibitem{land1977retinex}
E.~H. Land, ``{The retinex theory of color vision},'' \emph{Sci. Am.}, vol.
  237, no.~6, pp. 108--129, 1977.

\bibitem{mertens2009exposure}
\BIBentryALTinterwordspacing
T.~Mertens, J.~Kautz, and F.~{Van Reeth}, ``{Exposure Fusion: A Simple and
  Practical Alternative to High Dynamic Range Photography},'' \emph{Comput.
  Graph. Forum}, vol.~28, no.~1, pp. 161--171, Mar. 2009.
\BIBentrySTDinterwordspacing

\bibitem{harada1998multidimensional}
Y.~Harada, S.~Muramatsu, and H.~Kiya, ``{Multidimensional Multirate Filter
  without Checkerboard Effects},'' in \emph{Proc. Eur. Signal Process. Conf.},
  1998, pp. 1881--1884.

\bibitem{tamura1998design}
T.~Tamura, M.~Kato, T.~Yoshida, and A.~Nishihara, ``{Design of
  Checkerboard-Distortion-Free Multidimensional Multirate Filters},''
  \emph{IEICE Trans. Fundam. Electron. Commun. Comput. Sci.}, vol. E81-A,
  no.~8, pp. 1598--1606, Aug. 1998.

\bibitem{harada1998multidimensional_trans}
Y.~Harada, S.~Muramatsu, and H.~Kiya, ``{Multidimensional Multirate Filter and
  Filter Bank without Checkerboard Effect},'' \emph{IEICE Trans. Fundam.
  Electron. Commun. Comput. Sci.}, vol. E81-A, no.~8, pp. 1607--1615, Aug. 1998.

\bibitem{iwai2010methods}
\BIBentryALTinterwordspacing
H.~Iwai, M.~Iwahashi, and H.~Kiya, ``{Methods for Avoiding the Checkerboard
  Distortion Caused by Finite Word Length Error in Multirate System},''
  \emph{IEICE Trans. Fundam. Electron. Commun. Comput. Sci.}, vol. E93-A,
  no.~3, pp. 631--635, 2010.
\BIBentrySTDinterwordspacing

\bibitem{simonyan2014very}
\BIBentryALTinterwordspacing
K.~Simonyan and A.~Zisserman, ``{Very Deep Convolutional Networks for
  Large-Scale Image Recognition},'' \emph{arXiv: 1409.1556}, Sep.
  2014. [Online]. Available: \url{http://arxiv.org/abs/1409.1556}
\BIBentrySTDinterwordspacing

\bibitem{he2016deep}
\BIBentryALTinterwordspacing
K.~He, X.~Zhang, S.~Ren, and J.~Sun, ``{Deep Residual Learning for Image
  Recognition},'' in \emph{Proc. IEEE Conf. Comput. Vis. Pattern
  Recognit.},
  Jun. 2016, pp. 770--778.
\BIBentrySTDinterwordspacing

\bibitem{goodfellow2016deep}
I.~Goodfellow, Y.~Bengio, and A.~Courville, \emph{{Deep learning}},
  MIT Press, 2016.

\bibitem{marnerides2018expandnet}
\BIBentryALTinterwordspacing
D.~Marnerides, T.~Bashford-Rogers, J.~Hatchett, and K.~Debattista,
  ``{ExpandNet: A Deep Convolutional Neural Network for High Dynamic Range
  Expansion from Low Dynamic Range Content},'' in \emph{Comput. Graph. Forum},
  vol.~37, no.~2, Wiley Online Library,
  May 2018, pp. 37--49.
\BIBentrySTDinterwordspacing

\bibitem{kinoshita2019itmnet_trans}
\BIBentryALTinterwordspacing
Y.~Kinoshita and H.~Kiya, ``{iTM-Net: Deep Inverse Tone Mapping Using Novel
  Loss Function Considering Tone Mapping Operator},'' \emph{IEEE Access},
  vol.~7, pp. 73\,555--73\,563, 2019.
\BIBentrySTDinterwordspacing

\bibitem{ioffe2015batch}
\BIBentryALTinterwordspacing
S.~Ioffe and C.~Szegedy, ``{Batch Normalization: Accelerating Deep Network
  Training by Reducing Internal Covariate Shift},'' \emph{arXiv: 1502.03167},
  Feb. 2015. [Online]. Available:
  \url{http://arxiv.org/abs/1502.03167}
\BIBentrySTDinterwordspacing

\bibitem{glorot2011deep}
X.~Glorot, A.~Bordes, and Y.~Bengio, ``{Deep sparse rectifier neural
  networks},'' in \emph{Proc. Int. Conf. Artif. Intell. Stat.},
  Apr. 2011, pp. 315--323.

\bibitem{kingma2014adam}
\BIBentryALTinterwordspacing
D.~P. Kingma and J.~Ba, ``{Adam: A Method for Stochastic Optimization},''
  \emph{arXiv: 1412.6980}, Dec. 2014. [Online]. Available:
  \url{http://arxiv.org/abs/1412.6980}
\BIBentrySTDinterwordspacing

\bibitem{he2015delving}
\BIBentryALTinterwordspacing
K.~He, X.~Zhang, S.~Ren, and J.~Sun, ``{Delving Deep into Rectifiers:
  Surpassing Human-Level Performance on ImageNet Classification},'' in
  \emph{Proc. IEEE Int. Conf. Comput. Vis.},
  Dec. 2015, pp. 1026--1034.
\BIBentrySTDinterwordspacing

\bibitem{wang2004image}
\BIBentryALTinterwordspacing
Z.~Wang, A.~Bovik, H.~Sheikh, and E.~Simoncelli, ``{Image Quality Assessment:
  From Error Visibility to Structural Similarity},'' \emph{IEEE Trans. Image
  Process.}, vol.~13, no.~4, pp. 600--612, Apr. 2004.
\BIBentrySTDinterwordspacing

\bibitem{mittal2013making}
\BIBentryALTinterwordspacing
A.~Mittal, R.~Soundararajan, and A.~C. Bovik, ``{Making a “Completely
  Blind” Image Quality Analyzer},'' \emph{IEEE Signal Process. Lett.},
  vol.~20, no.~3, pp. 209--212, Mar. 2013.
\BIBentrySTDinterwordspacing

\bibitem{kinoshita2018automatic}
\BIBentryALTinterwordspacing
Y.~Kinoshita, S.~Shiota, and H.~Kiya, ``{Automatic Exposure Compensation for
  Multi-Exposure Image Fusion},'' in \emph{Proc. IEEE Int. Conf. Image
  Process.}, Oct. 2018, pp.
  883--887.
\BIBentrySTDinterwordspacing

\end{thebibliography}

\end{document}